\documentstyle[twocolumn,prl,aps,epsfig]{revtex}
\begin{document} 
\title{The Edge-State Theory of Integer-Quantum-Hall-Effect to Insulator Transition}
\author{X. R. Wang$^1$, X. C. Xie$^2$, Q. Niu$^3$, and J. K. Jain$^4$}
\address{
$^1$ Department of Physics, The Hong Kong University of Science
and Technology, Clear Water Bay, Hong Kong}
\address{
$^2$ Department of Physics, Oklahoma State University, Stillwater, 
OK 74078}
\address{
$^3$ Department of Physics, University of Texas, Austin, Texas 78712}
\address{
$^4$ Department of Physics, Pennsylvania State University, University 
Park, PA 16802
}

\address{\mbox{ }}
\address{\parbox{14cm}{\rm \mbox{ }\mbox{ }
Direct transitions, driven by disorder, from several
integral quantum Hall states to an insulator
have been observed in experiment.  This finding is enigmatic 
in light of a theoretical phase diagram, based on rather general 
considerations, that predicts a sequence of transitions in which the integer $n$ 
characterizing 
the Hall conductivity is reduced successively by unity, eventually going from $n=1$ into 
an insulator.  In this work, we suggest that the direct transition occurs because,
in certain parameter regime, the edge states of different Landau levels are 
strongly coupled and behave as a single edge state.
It is indicated under what conditions successive transitions may be seen.
}}
\address{\mbox{ }}
\address{\parbox{14cm}{\rm PACS numbers: 73.40.Hm, 71.30.+h, 73.20.Jc}}
\maketitle


\vspace{-0.5cm}

The conventional scaling theory of localization\cite{scaling} predicts
that all electrons in a two-dimensional system are localized in the 
absence of a magnetic field. In the presence of a strong magnetic field, 
the energy spectrum becomes a series of disorder broadened Landau bands.
The phenomenon of the integral quantum Hall effect (IQHE) indicates 
the existence of extended states at the center of each Landau band, separated
by localized states at other energies. The integrally quantized plateaus are  
observed when the Fermi level lies in the localized states, with the 
value of the Hall resistance, $R_H=h/ne^2$, related to the number of extended 
bands ($n$) below the Fermi energy.  As a function of the magnetic field, 
the Hall resistance jumps from one quantized value to another when the Fermi 
energy crosses an extended-state level.

In order to reconcile the presence of extended states at finite magnetic fields
with the lack thereof at zero magnetic field, Khmelnitskii\cite{khme} and
Laughlin\cite{laughlin} came up with a picture in which the extended states associated 
with Landau levels levitate due to disorder, eventually pushed to very high
energies, leaving an Anderson insulator behind.  Based on this physics,
Kivelson, Lee and Zhang\cite{kivelson} (KLZ) proposed 
a topological phase diagram as shown in 
Fig.~1(a).   A crucial prediction of this phase diagram is that an IQHE 
state $n$ in general can only go into another IQHE states $n\pm 1$, and that 
a transition into an insulating state is allowed only from the $n=1$ state.

This phase diagram motivated numerous theoretical and experimental studies.
\cite {dzliu,yang,sheng,haldane,jiang,tkwang,glozman,kravchenko,song}.
However, direct transitions 
from $n=1,\ n=2,\ n=3,$ and $n=6$ IQHE states to the insulator 
have been observed in recent experiments\cite{kravchenko,song}. 
The transition from $n=2$ IQHE state to the Anderson insulator may 
perhaps be related to the spin degeneracy in the lowest Landau level\cite
{fogler}, but the other transitions are inconsistent with the KLZ phase diagram.

\begin{figure}
\vspace{2mm}
\centering
\epsfig{file=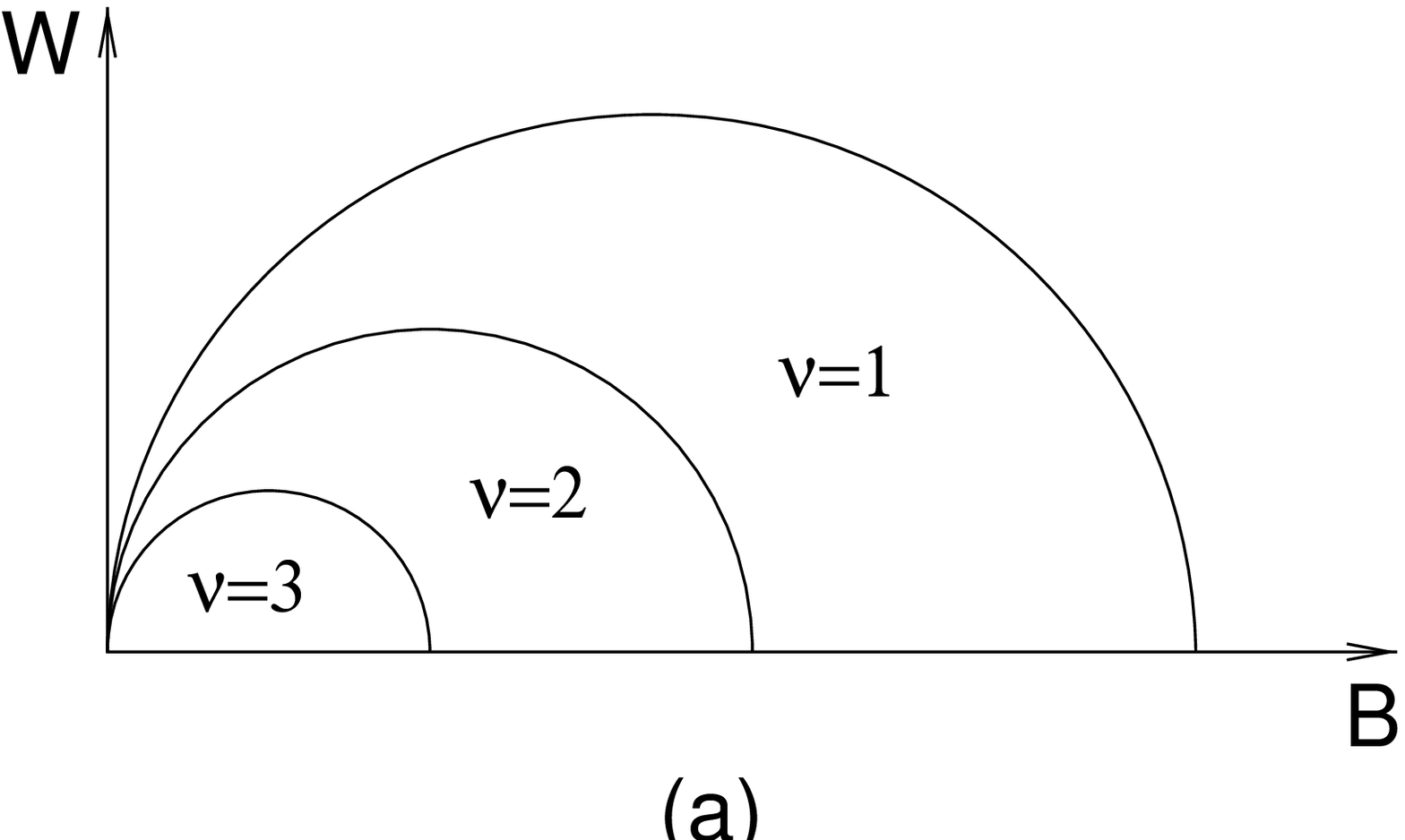, width=0.8\columnwidth}
\vskip 0.5cm
\epsfig{file=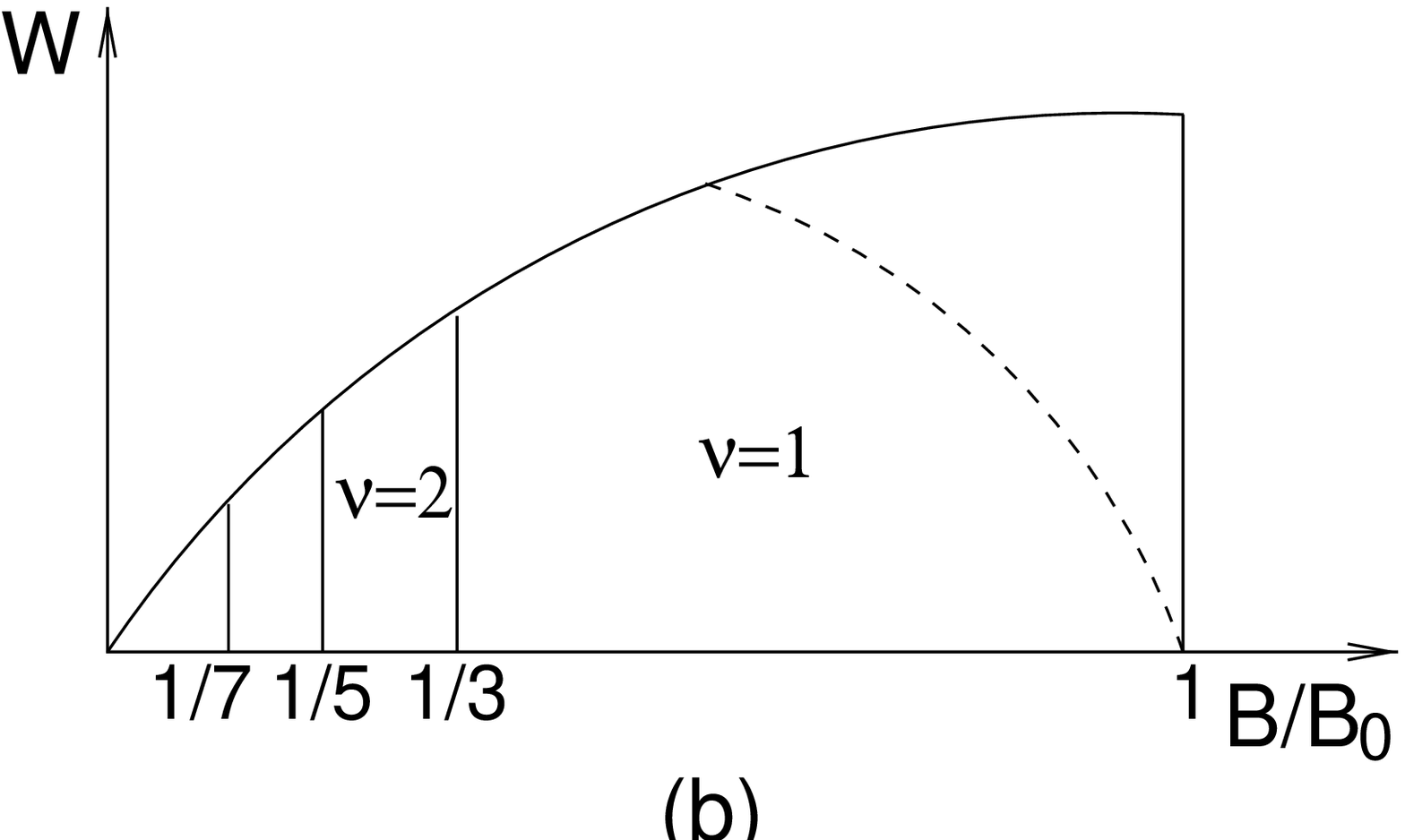, width=0.8\columnwidth}
\vskip 0.5cm
\caption{(a) The nested phase diagram of Kivelson, Lee, and 
Zhang.\protect\cite{kivelson} Its topology does not allow direct 
transitions form an $n\neq 1$ IQHE state to an Anderson 
insulator. The only allowed transitions are $n\rightarrow n\pm 1$.
(b) The non-nested phase diagram as originally predicted by 
Liu, Xie, and Niu\protect\cite{dzliu} (solid line) and confirmed and
modified by Sheng and Weng\protect\cite{sheng} (dashed line).
$B_{0}$ is the field when the lowest Landau level is half-occupied. 
\label{fig1}}
\end{figure} 

Recently, Liu, Xie and Niu\cite{dzliu} studied numerically a 
tight-binding model of two-dimensional electrons in a magnetic 
field and a random potential. They found that 
the extended levels do not float up
to infinity but are instead destroyed by strong disorder at a critical 
magnetic field. They calculated phase diagrams in the energy-field
and energy-disorder planes, which may be combined to yield
a non-nested phase diagram shown in Fig.~1(b) (solid line).
In this phase diagram transitions from any 
IQHE state to the insulating phase are allowed when the disorder 
is increased at fixed B. The non-nested 
phase diagram of Fig.~1(b) was confirmed by recent calculations with an
important modification on the lowest Landau level\cite{sheng}
(dashed line), which is now in excellent agreement with
experiments\cite{kravchenko,song}.

Thus far, none of the theoretical calculations in this context includes 
explicitly the edge states, whose consideration is crucial for transport 
in the QHE regime for the following reasons.  First, 
as clarified by the Landauer approach, \cite{Jain} and 
confirmed in numerous experiments, the Hall current is 
carried to a large extend by the edge states.   
On the QHE plateaus, backscattering is exponentially suppressed due to a spatial separation 
of edge states carrying current in opposite directions, which is responsible 
for the spectacular accuracy of the quantization of the Hall resistance. 
A non-zero longitudinal resistance owes its origin to 
an inter-edge backscattering.   An insulator is obtained when the edge channels are 
fully reflected backwards, with the  
forward transmission coefficient becoming exponentially small.  Thus, an IQHE to insulator
transition occurs when the edge channels shut off.
Furthermore, Landau level mixing due to disorder, relevant for the issue at hand, 
is expected to be particularly strong at the edges, 
because here the states from different LLs are close not only in space but also in 
energy.   The goal of the present work is to investigate the mixing of 
edge states on the same edge of the sample relative to the mixing of 
edge states on opposite edges, responsible for inter-edge scattering.

We will work within a continuum model.
A numerical study of a two-dimensional continuum model
is difficult due to the limitation in computational power.
Therefore, we study a narrow strip and ask: 
How does a quantum Hall state
evolves as the magnetic field decreases or as disorder increases?

To be specific, we consider the system shown in Fig.~2,
in which there are two edge states at each edge of the sample.  
The states on a given edge of the sample are ``chiral", i.e., 
carry current only 
in one direction, determined by the $\vec{E}\times\vec{B}$ drift, with 
the electric field provided by the confinement potential. 
The edge states carry current in opposite directions on the opposite 
sides of the sample. Each edge
carries a current of $\frac{|e|} {\hbar}$ per unit energy.
The disorder causes couplings between the various edge states.
The strength of the coupling depends on the overlap of the states,
which increases with disorder.  Coupling of the edge states on the
same side by disorder  gives only 
forward-scattering, which does not degrade the source-to-drain current.
The only effect of such a scattering is an unimportant
forward-scattering phase shift\cite{dassarma}. The current can be
diminished only by scattering 
between two edge states on the opposite sides -- namely, backscattering.
When the reflection coefficient becomes unity, the 
edge states shut off, no current flows through, and an insulator is obtained.

Clearly, an understanding the
transition from QHE states to the insulator is intimately related to 
the evolution of the edge states as a function of disorder.
Two different scenarios are possible. 
(i)  The edge channels shut off one by one.
This would correspond to the nested KLZ phase diagram.  
(ii) The edge channels shut off
all at once.  This would correspond to the non-nested phase diagram 
of Fig.~1(b).   
Which scenario occurs depends on the relative strengths of the couplings of 
edge channels on the same side and of edge channels on the opposite sides. 
The former scenario is likely when the coupling between the edges states 
on the same side is weak, whereas the latter scenario is expected to occur when
the coupling between the edge states on the same side is strong. 
This is the physical picture that we propose as an explanation of 
the direct transition between high integer IQHE states and an insulator. 
In the remainder of the paper, we confirm this through an explicit calculation.

\begin{figure}
\vspace{2mm}
\centering
\epsfig{file=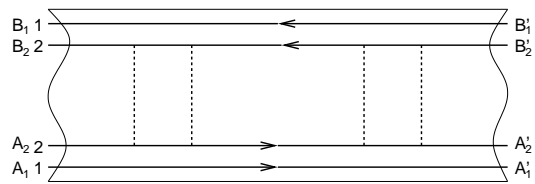, width=0.8\columnwidth}
\vskip 0.5cm
\caption{
Two edge channels in a sample. The channels on the top carry current 
from right to left, and those at 
the bottom carry the current in the opposite direction, as indicated by 
arrows.  The backscattering is indicated by dashed lines. 
\label{fig2}}
\end{figure}


In Fig.~2, edge states 1 and 2 at bottom 
carry current from left to right, and 
edge states 1 and 2 on the top carry current in the 
opposite direction. Electrons can move from one side of the system 
to the other side through the four channels provided by the two 
edge states. Disorders may scatter electrons from one channel
into another. Assume that $A_{1}, \  A_{2}$ and $B_{1}', \ 
B_{2}'$ are the amplitudes of wavefunction in edge states 1 and 2 
going into the sample from the left-hand and right-hand sides of the 
system, respectively, and similarly, that $B_{1}, \  B_{2}$ and 
$A_{1}',\  A_{2}'$ are the amplitudes of wavefunction in edge states 
1 and 2  coming out of the sample from the left-hand and 
right-hand sides of the system, respectively.  These amplitudes are 
related by an S-matrix  
\begin{equation}
\label{smatrix}
\left(\begin{array}{c}B_{1}\\B_{2}\\A_{2}'\\A_{1}' \end{array} \right)=
S\left(\begin{array}{c}B_{1}'\\B_{2}'\\A_{2}\\A_{1} \end{array} \right),
\end{equation}
or an M-matrix
\begin{equation}
\label{tmatrix}
\left(\begin{array}{c}B_{1}'\\B_{2}'\\A_{2}'\\A_{1}' \end{array} \right)=
M\left(\begin{array}{c}B_{1}\\B_{2}\\A_{2}\\A_{1} \end{array} \right),
\end{equation}
where 
\begin{equation}
\label{smatrix1}
S=\left(\begin{array}{cccc}t_{11}'&t_{21}'&r_{21}&r_{11}\\
t_{12}'&t_{22}'&r_{22}&r_{12}\\
r_{12}'&r_{22}'&t_{22}&t_{12}\\
r_{11}'&r_{21}'&t_{21}&t_{11}\\
\end{array} \right),
\end{equation}
and the matrix $M$ can be obtained from $S$. 
Elements $t_{ij}$ and $t_{ij}'$ are 
the transmission coefficients of electron moving
from edge state $i$ to state $j$ from left-hand side and right-hand side of
the system, respectively, while $r_{ij}$ and $r_{ij}'$ are the 
reflection coefficients of electron moving from edge state $i$ to state 
$j$ in the left-hand side and right-hand side of the system, respectively.
The probability conservation requires $S$ to be unitary. $M$ will be  
useful for the numerical calculations.

In order to model disorder, we introduce
backscattering and forward-scattering of edge channels 1 and 2, 
characterized by $r_{ij}$ and $t_{ij} \ (i \ne j)$.
To resolve the question of whether the two channels 
switch off simultaneously or 
one by one, we define two localization lengths, $\xi_1$ and $\xi_2$.
Assume that electrons have the same probability in both channels 1 and 2
at far left-hand side of the system.  Then the probabilities to find
electrons in channels 1 and 2 at right-hand side are proportional to
$|t_{11}|^{2}+|t_{21}|^{2}$ and $|t_{12}|^{2}+|t_{22}|^{2}$, respectively.
The localization lengths of channel 1 and channel 2 are then given by
\begin{equation}
\label{length1}
\xi_{1}^{-1}=-\lim_{L\rightarrow \infty }\frac{\ln(|t_{11}|^{2}+|t_{21}|^{2})}
{ 2 L},
\end{equation}
\begin{equation}
\label{length2}
\xi_{2}^{-1}=-\lim_{L\rightarrow \infty }\frac{\ln(|t_{12}|^{2}+|t_{22}|^{2})}
{2 L},
\end{equation}
where $L$ is the size of the system.
The ratio $\xi_1/\xi_2$ vanishes when the two edge channels switch off 
separately (with channel 1 switching off first), and becomes unity when 
the two edge channels shut off at the same time. 

\begin{figure}
\vspace{2mm}
\epsfig{file=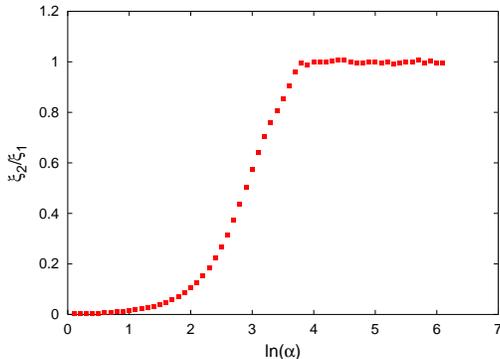, width=0.8\columnwidth}
\caption{$\frac{\xi_{2}} {\xi_{1}}$ vs $\ln \alpha$. 
$\xi_{1}$ and $\xi_2$ are the localization lengths of channels
1 and 2, respectively. $\alpha$ is the ratio of the backward scattering 
strength of channel 2 and forward scattering between channels 1 and 2.
\label{fig3}}
\end{figure}


The localization lengths of the two channels can be calculated 
by using the multiplicative property of the M-matrix.  We first divide the 
system into many vertical slices.  Each slice contains only very small number of 
disorders so that the partial $S$ matrix for each slice is close to the 
unit matrix.  The backscattering is expected to be the strongest for the innermost 
edge channel.  For simplicity, we assume that backward-scattering occurs  
only through edge channel 2, i.e. set $r_{11}=0= r_{11}'$.
We further assume that randomness comes from the phase of each scattering 
process rather than its amplitude. Under these assumptions, the general
unitary matrix $S$ becomes
\[
S=\left(\begin{array}{cccc}\sqrt{1-t^{2}}&t e^{i\phi_{1}}&0&0\\
-t e^{-i\phi_{1}}&\sqrt{1-t^{2}}&0&0\\
0&0&\sqrt{1-t^{2}}&t e^{i\phi_{2}}\\
0&0&-t e^{-i\phi_{2}} &\sqrt{1-t^{2}}\\
\end{array} \right) \times 
\]
\begin{equation}
\label{Amatrix}
\left(\begin{array}{cccc}1&0&0&0\\
0&\sqrt{1-r^{2}}&re^{i\phi_{3}}&0\\
0&-r e^{-i\phi_{3}}&\sqrt{1-r^{2}}&0\\
0&0&0&1\\
\end{array} \right),
\end{equation}
where $r$ measures the strength of backward-scattering in channel 2, and 
$t$ is the strength of forward-scattering from channel 1 to channel 2, 
and vice versa. The quantities $t$ and $r$ are chosen to be small 
in order to ensure  
that scattering in each slice is weak. Throughout this study, we set $r=0.0002$ and 
vary $t$. The ratio $\alpha=t/r$ measures the relative strength of 
backscattering and forward-scattering.
$\phi_{1}, \ \phi_{2},$ and $\phi_{3}$ are random numbers. 
The random matrix determines the $S$ matrix of the system with
disordered thin slice, 
which, in turn, gives a random $M$ matrix. The product of 
$L$ independent random matrices of this type is used to calculate 
localization lengths of edge states \cite {xie}.
The numerical result of $\frac{\xi_{2}} {\xi_{1}}$ vs 
$\ln \alpha$ is plotted in Fig.~3.

The principal result is that $\frac{\xi_{2}} {\xi_{1}}$ changes 
from 0 to 1 when $\ln \alpha$ increases from -1 to 4.
$\frac{\xi_{2}} {\xi_{1}}=0$ means that channel 2 shuts off  
when channel 1 is still propagating from source to drain (or 
vice versa).  This confirms that for small $\alpha$, the edge channels are 
blocked off one by one through 
backscattering.  In this case, transitions of IQHE states to the insulator 
should be governed by the phase diagram of KLZ.
On the other hand, for large $\alpha$, both edge channels shut off 
at the same time.  It is remarkable that this happens even though 
the backscattering occurs in our model only in channel 2.
This implies a direct transition from an arbitrary IQHE state to an insulator, 
consistent with the non-nested phase diagram of Fig.~1(b).

We would like to make several remarks. 1) In our approach, we only
consider phase randomness due to disorder scattering.  As usual, the amplitude
randomness is not expected change the results. The structure of random
matrix M is the same whether only phase randomness or both phase
and amplitude randomness are taken into account. 2) 
The parameter $\alpha$ depends on the disorder potential, $V(r)$, as well as the 
sample width, $W$.  Both forward and backward scatterings 
are proportional to the matrix element of the disorder 
potential between the two states involved, $<\Psi_{1}|V|\Psi_{2}>$. 
When $\Psi_{1}$ and $\Psi_{2}$ are the edge states on the same side of
a sample, the matrix element, which characterizes the forward scattering, 
is proportional to $\exp (-d^{2}/l_{B}^{2})$, where  the distance $d$ between 
the centers of the two edge states is on the order of the magnetic length, 
$l_{B}=\sqrt{\hbar/eB}$.  The gaussian expression originates from the magnetic 
confinement on the Landau level wavefunction.  On the other hand, when $\Psi_{1}$ and 
$\Psi_{2}$ correspond to the edge states on the opposite sides of the sample, 
the matrix element for backscattering is of order $\exp (-W/\xi_t)$, where $\xi_t$ is 
a disorder dependent length characterizing the transverse extent of the edge state.
(Even though the electronic wave function of a perfect system in a magnetic field 
is a gaussian, it has an exponential tail in the presence of disorder.\cite{Thouless})
This exponential suppression of backscattering as a function of the width suggests that
sufficiently wide samples are in the $\ln \alpha>4$ regime, governed by the 
non-nested phase diagram of Fig.~1b.  
Narrow samples, however, may be in the opposite regime.
The actual width at which the crossover occurs 
depends on the disorder potential,
but may be expected to be on the order of several magnetic lengths. 


According to the picture presented here, a direct transition from an arbitrary
IQHE state to an insulating state is intimately connected to the Landau level mixing at the 
edges of the sample.  At first sight, this physics appears to bear no relation 
to the direct transitions 
seen in earlier numerical studies, because these studies employed  
periodic boundary conditions, 
i.e., the numerical samples had no edges due to a geometrical confinement.  
However,  we believe that the underlying physics may be identical.  Even 
though there are no real edges, internal edges, corresponding to equipotential 
contours, appear in the presence of disorder.  When the disorder is sufficiently 
strong, the internal edges carry, in general, edge channels from several LL's. 
The IQHE to insulator transition is a percolation transition in this picture, 
which has to do with the coupling between the network of internal edges.  
Again, if the channels from different LL's are strongly coupled and behave 
effectively as a single edge channel, there would be a direct transition 
from IQHE to insulator.

In summary, we have investigated the role of edge states in disorder driven  
IQHE to insulator transition.  We find that a direct transition from a general 
IQHE state to an insulator is possible when the edge channels of different 
Landau levels are strongly coupled.

X.R.W. would like to thank Mr. Xiong Gang for numerical assistance.
We acknowledge support from UGC, Hong Kong, through RGC grant (X.R.W.), 
US DOE-98ER45687 and DOE-99ER45755(X.C.X.),
NSF-DMR9705406, the
Welch Foundation and China-NSF (Q.N),
and NSF DMR-9986806 (J.K.J.).


\end{document}